\documentclass[a4paper, 10pt,  conference]{IEEEtran}
\IEEEoverridecommandlockouts
\makeatletter
\newcommand{\linebreakand}{%
  \end{@IEEEauthorhalign}
  \hfill\mbox{}\par
  \mbox{}\hfill\begin{@IEEEauthorhalign}
}
\makeatother

\usepackage{cite}
\usepackage{amsmath,amssymb,amsfonts}
\usepackage{algorithm}
\usepackage{amsmath}
\usepackage{algpseudocode}
\usepackage{graphicx}
\usepackage{textcomp}
\usepackage{xcolor}
\def\BibTeX{{\rm B\kern-.05em{\sc i\kern-.025em b}\kern-.08em
    T\kern-.1667em\lower.7ex\hbox{E}\kern-.125emX}}
\begin{document}

\title{Vacuum Circuit Breaker Closing Time Key Moments Detection via Vibration Monitoring: \\A Run-to-Failure Study\\
\thanks{The presented work is part of a project that is financially supported by the Swiss Federal Office of Energy.}
}

\author{\IEEEauthorblockN{1\textsuperscript{st} Chi-Ching Hsu}
\IEEEauthorblockA{\textit{High Voltage Laboratory} \\
\textit{ETH Zurich}\\
Zurich, Switzerland \\
hsu@eeh.ee.ethz.ch}
\and
\IEEEauthorblockN{2\textsuperscript{nd} Gaetan Frusque}
\IEEEauthorblockA{\textit{ Intelligent Maintenance and Operations Systems Laboratory} \\
\textit{EPFL}\\
Lausanne, Switzerland \\
gaetan.frusque@epfl.ch}
\and
\IEEEauthorblockN{3\textsuperscript{rd} Mahir Muratovic}
\IEEEauthorblockA{\textit{High Voltage Laboratory} \\
\textit{ETH Zurich}\\
Zurich, Switzerland \\
muratovic@eeh.ee.ethz.ch}
\linebreakand
\IEEEauthorblockN{4\textsuperscript{th} Christian M. Franck}
\IEEEauthorblockA{\textit{High Voltage Laboratory} \\
\textit{ETH Zurich}\\
Zurich, Switzerland \\
franck@eeh.ee.ethz.ch}
\and
\IEEEauthorblockN{5\textsuperscript{th} Olga Fink}
\IEEEauthorblockA{\textit{Intelligent Maintenance and Operations Systems Laboratory} \\
\textit{EPFL}\\
Lausanne, Switzerland \\
olga.fink@epfl.ch}
}

\maketitle

\begin{abstract}
Circuit breakers (CBs) play an important role in modern society because they make the power transmission and distribution systems reliable and resilient. Therefore, it is important to maintain their reliability and to monitor their operation. A key to ensure a reliable operation of CBs is to monitor their condition. In this work, we performed an accelerated life testing for mechanical failures of a vacuum circuit breaker (VCB) by performing close-open operations continuously until failure. We recorded data for each operation and made the  collected run-to-failure dataset publicly available. In our experiments, the VCB operated more than 26000 close-open operations without current load with the time span of five months. The run-to-failure long-term monitoring enables us to monitor the evolution of the VCB condition and the degradation over time. To monitor CB condition, closing time is one of the indicators, which is usually measured when the CB is taken out of operation and is completely disconnected from the network. We propose an algorithm that enables to infer the same information on the closing time from a non-intrusive sensor. By utilizing the short-time energy (STE) of the vibration signal, it is possible to identify the key moments when specific events happen including the time when the latch starts to move, and the closing time. The effectiveness of the proposed algorithm is evaluated on the VCB dataset and is also compared to the binary segmentation (BS) change point detection algorithm. This research highlights the potential for continuous online condition monitoring, which is the basis for applying future predictive maintenance strategies.





\end{abstract}


\section{Introduction}
\label{sec:intro}
Electricity is indispensable in modern society and circuit breakers (CBs) play a vital role in making the power transmission and distribution systems highly reliable and resilient. According to IEEE standard C37.100-1992, "CB is a mechanical switching device, capable of making, carrying and breaking currents under normal circuit conditions and also, making and carrying for a specified time and breaking currents under specified abnormal circuit conditions such as those of short circuit"~\cite{ieee1992ieee}. They are well known for their long lifespan of up to several decades and low failure rates of about 0.30 to 1.58 failures per 100 CB years~\cite{janssen2013international, Carvalho2012CIGRTBP1}. While they have long useful lifetimes, they only operate when abnormal conditions appear, which results in rare operations. For CBs, any unexpected failure leads to potentially catastrophic consequences. As a result, making sure that CBs operate smoothly without failure is a key to maintaining highly reliable power systems. 

Common CB failures include mechanical and electrical parts. According to the CIGRE survey conducted in the year 2004 to 2007 from 26 countries and 83 utilities~\cite{razi2020condition, Carvalho2012CIGRTBP1}, mechanical parts such as operating mechanisms, compressors, pumps and actuators are reported to be one of the main causes of all occurring faults with 43.5\%, followed by electrical parts such as high voltage parts with a share of 26\%, and other parts such as control and auxiliary components including coil current, auxiliary switch faults with 24.5\%.

One conventional method of monitoring the CB condition is measuring the mechanical opening and closing times. With the degradation, both opening and closing times increase~\cite{razi2015applicability, razi2020condition}. These measurements reflect the CB condition and are based on the movement of the CB contacts. Contacts are the conducting part in CB that moves to make or break a circuit. Opening time is defined as the time interval between the time when the actuating signal is sent, and the instant when the contacts have parted. Closing time is similarly defined as the time interval between the time when the actuating signal is sent, and the instant when the contacts have touched~\cite{ieee1992ieee}. 

There are two common approaches to measure opening and closing times: travel curve and contact separation measurement. However, both approaches face the same problem that CB needs to be taken out of service from the grid in order to perform the measurements~\cite{razi2015applicability}. These two approaches can, therefore, not be used for continuous online monitoring. The first approach, travel curve (also referred as motion curve), uses motion sensors such as a transducer or travel encoder to record contact position. With it, it is possible to infer not only opening and closing times but also opening and closing speeds and accelerations. However, a travel curve sensor is not easy to install routinely and reliably on every CB in the field. The second approach, contact separation measurement, is measured by applying a low voltage between the contacts and measuring the voltage difference. When the contacts open, the voltage between the contacts goes from zero to a finite low voltage signal. Similarly to the inference of closing time, we can determine the contact touch during close operation when the voltage signal goes from the finite low voltage to zero. The start time when measuring opening and closing time is defined as the time when an operating signal is sent to the corresponding coil. The opening and closing time measurements are precise by this approach, because it infers directly when contacts are touched and separated. However, the CB still needs to be disconnected from the grid.


In this research, we performed an accelerated life testing for mechanical failures of a VCB by performing close-open operations continuously until failure. The condition of the VCB was monitored by the vibration measurements and also the contact separation measurements. 
The measurements of the vibration are performed non-intrusively since the vibration sensors can be installed outside the breaker chamber on the drive enclosure. As mentioned before, closing time can be directly used as a health indicator and can be intuitively understood by domain experts. While vibration signals contain a lot of information on the VCB condition and are non-intrusive, the interpretation of the extracted features by domain experts is more difficult and less intuitive. Therefore, extracting the same information from the vibration signals as has been used in the past by domain experts, will provide an easy way to interpret and monitor the VCB condition.


We propose a method to detect key moments in the closing process using vibration signal. These key moments are time points when specific events happen. With this non-intrusive condition monitoring approach, it becomes possible to infer key moments that distinguish different segments of the closing process without the need to disconnect the VCB from the grid. Identifying key moments allows us to better understand the VCB close operation. We compare the performance of detecting closing time from the vibration signals from our proposed method to several change point detection algorithms including binary segmentation (BS), window sliding, and bottom-up segmentation.






In addition, to understand the ageing behavior over the entire lifetime and the potential failure mechanisms of a CB, we performed an accelerated mechanical life testing experiment on a VCB without load. One of the goals is to be able to determine the remaining useful life (RUL). We collected a continuously monitored run-to-failure dataset without any artificially induced faults. We monitored the VCB with both intrusive and non-intrusive sensors and recorded the corresponding data. To the best of our knowledge, this is the first publicly available run-to-failure VCB dataset with more than 26000 open and close operations. Recorded signals include coil current, opening and closing times using contact separation measurements, vibration signals, and motor charging current for every open and close operation. 



The paper is organized as follows: first, the related works are reviewed in Section~\ref{sec:literature}. The experimental setup is explained in Section~\ref{sec:experiment}. The proposed key moments detection method is described in Section~\ref{sec:methods} and results on the VCB dataset are presented in Section~\ref{sec:results}. Finally, the results are discussed in Section~\ref{sec:discussion} and final conclusions are provided in Section~\ref{sec:conclusions}.


\section{Related works}
\label{sec:literature}
Different condition monitoring approaches have been proposed to monitor the condition of CBs. The most commonly used condition monitoring signals include vibration signals~\cite{meng2006detection, hoidalen2005continuous, yang2019chaotic, dou2018application, yang2019new, ma2018intelligent, ukil2013monitoring}, coil current~\cite{johal2008coil, razi2016circuit}, dynamic resistance measurements~\cite{liu2018prediction}, opening and closing times~\cite{razi2015applicability, rusek2008timings}. Vibration monitoring has been the most frequently used condition monitoring approach for CBs in the literature in the last 20 years and coil current has been the second~\cite{razi2020condition}. In this section, we focus only on the condition monitoring signals applied in this study, namely, vibration signals and opening and closing time measurement.  

\textbf{Vibration} signals during open and close operations are vital for CB condition monitoring. A comprehensive vibration analysis on three spring-operated SF\textsubscript{6} CBs from different manufacturers under three years of continuous condition monitoring are presented in the work~\cite{hoidalen2005continuous}. In total, more than 1000 vibration patterns are collected with accelerometers for assessing the CB condition. This work compares vibration patterns with a reference pattern to find a deviation in signals. The CB condition is assessed based on these deviations. Therefore, domain knowledge about the reference patterns is needed for this method. It shows that detecting mechanical malfunctions in operating mechanisms is possible through continuous monitoring of vibration signals. 

Most of the studies on vibration monitoring have focused on fault detection and identification of different fault types, whereby the faults are usually introduced artificially ~\cite{ukil2013monitoring, hoidalen2005continuous, yang2019chaotic, dou2018application, yang2019new, ma2018intelligent}. One study~\cite{meng2006detection} has a similar focus as ours. By using short-time energy (STE) and signal-to-noise ratio of the VCB vibration signals, the detection of closing time under load and no-load conditions was investigated. It is possible to detect the closing time from the vibration signals based on a predefined threshold. Instead of only performing closing time detection, we detect two key moments and monitor the VCB condition over the entire VCB lifetime.

\textbf{Change point detection} algorithms detect the change point and split signals into different segments based on their characteristics. The change points are the points where the data distributions before and after the point are different. An extensive survey on change point detection can be found in~\cite{truong2020selective}. There are three key elements of change point detection methodology: cost functions, search methods and constraints. Cost functions characterise the homogeneity of the signals. It should be low if the signal is homogeneous and high otherwise. In terms of search methods, there are optimal and approximate methods. Due to the high computational complexity of optimal methods to find the exact solution, we focus on approximate methods here, which find only approximate solution. The last element constraints tackle with the problem if the number of change point is known.

One of the commonly used approximate methods is the window sliding method. It detects the change point by finding the discrepancy of the cost function between the neighboring windows along the signals. One of the other commonly applied approaches, the bottom-up segmentation first splits signals into small segments where all split points are potential change points. Based on the discrepancy of the cost function between adjacent segments, the point is removed if the discrepancy is small or kept if it is large. In the end, the points left are the change points. In addition, BS is similar to bottom-up segmentation. It finds the change point by finding the minimum of the sum of the cost function before and after the change point. More details on BS are provided in Section~\ref{sec:change_point}.

Change point detection algorithms are used in various fields such as speech processing and gait analysis. In the application of CB, it was implemented as online signature analysis for coil current signatures~\cite{johal2008coil}. In our case, we use a change point detection algorithm to detect the closing time from the vibration signals because the  distributions of the vibration signals before and after the contacts touched are different. 

\textbf{Run-to-failure degradation trajectories} with either continuous or regular condition monitoring are key for data-driven approaches to understand complicated systems. However, run-to-failure datasets are usually missing. Few public run-to-failure datasets are available in different domains such as bearings~\cite{wang2018hybrid} and aircraft engines~\cite{arias2021aircraft}. To the best of our knowledge, there is no run-to-failure VCB dataset publicly available. One study analyzes vibration signals without load in a run-to-failure VCB experiment~\cite{obarcanin2021} and shows the evolution of the time domain features extracted from vibration signals. However, in that study closing time was not analysed. Other than this, most of the previous works in literature have not focused on monitoring the evolution of the CB condition in time but rather focused on detecting and diagnosing artificially induced faults~\cite{ukil2013monitoring, hoidalen2005continuous, yang2019chaotic, dou2018application, yang2019new, ma2018intelligent}.

\section{Experimental Setup}
\label{sec:experiment}
\begin{figure}[htbp]
\centerline{\includegraphics[width=60mm]{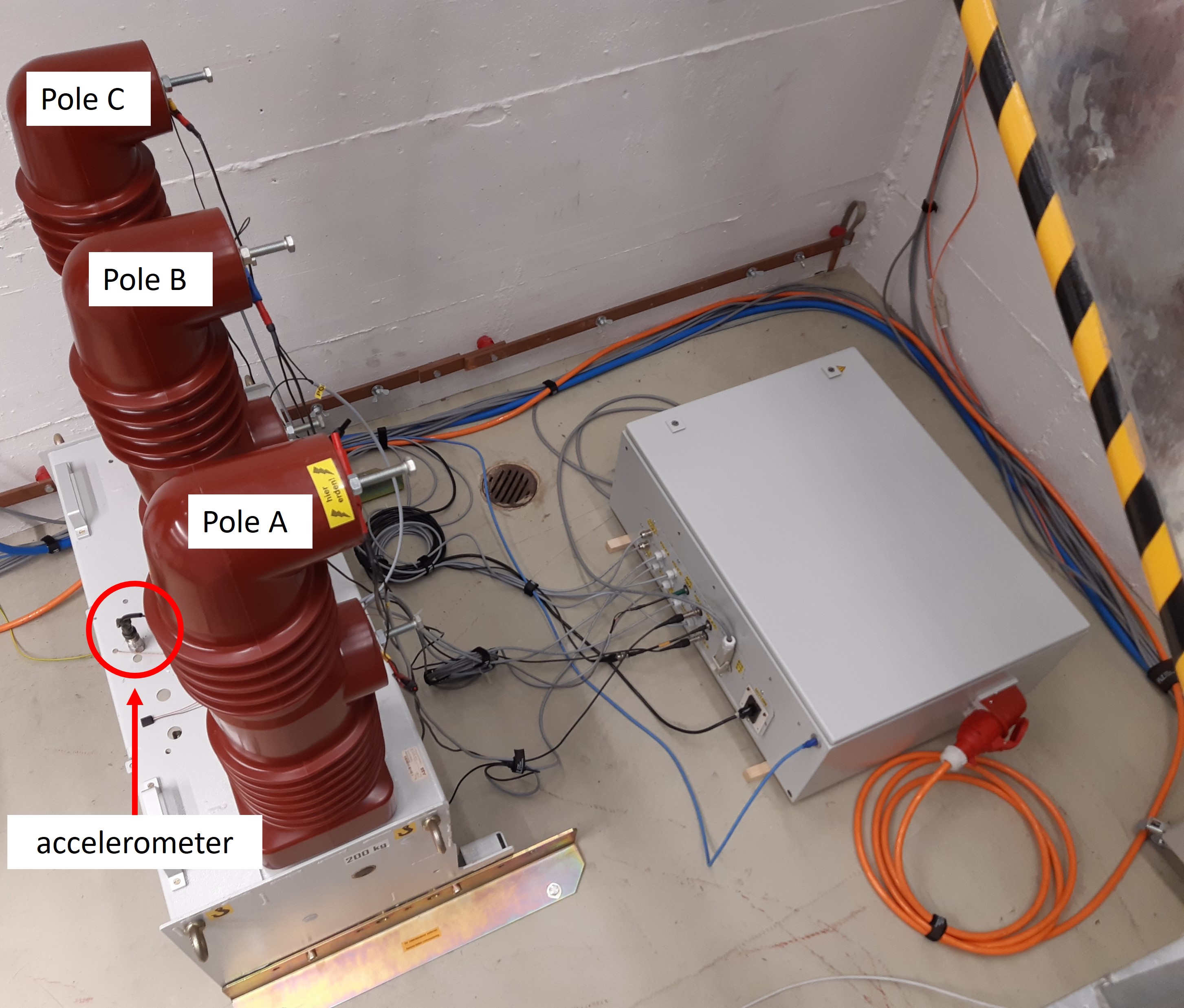}}
\caption{Experiment setup in the laboratory with the VCB on the left. Three poles are arbitrarily named as pole A, B, and C. The accelerometer is installed on the drive enclosure.}
\label{fig:experiment_setup}
\end{figure}

The test object is ABB VD4 Type 3612-16 spring-operated VCB with rated voltage of 36kV built in 2013 intended for indoor installation in air-insulated switchgear. The spring is charged via a 230V alternating current (AC) motor with a charging time of approximately 10s. The VCB can be operated manually by pressing the open and close buttons in the front panel and also by sending electric signals to the open and close coil.  Figure~\ref{fig:experiment_setup} shows the test VCB in the laboratory with three poles arbitrarily named as pole A, B, and C.

In this work, we perform a run-to-failure experiment on the VCB without current load and without any preventive maintenance in between. The experiment starts when the spring is fully discharged and the VCB is in its open position. A single close and open operation is performed in sequence of close-30s-open-3min. The experiment is controlled by the device LabJack T7 and data are recorded using TiePie digital oscilloscope. The data are recorded with sampling frequency of 300kHz and length of 200ms.


\subsection{Vacuum Circuit Breaker Dataset}
The test VCB has already performed 2442 open-close operations when the experiment started. The operation number is defined as the number of performed open operations and is shown on the counter on the front panel. The VCB experiences a fatal failure, which is caused by friction in the operating mechanism after around five months of continuous operations. In total, 26243 open and close operations are performed. For each operation, we recorded open and close coil current, opening and closing time, vibration signals, and motor charging current. The dataset is available to public on the ETH Zurich Research Collection~\cite{vcb_dataset2022}.

\begin{figure}[htbp]
\centerline{\includegraphics[width=60mm]{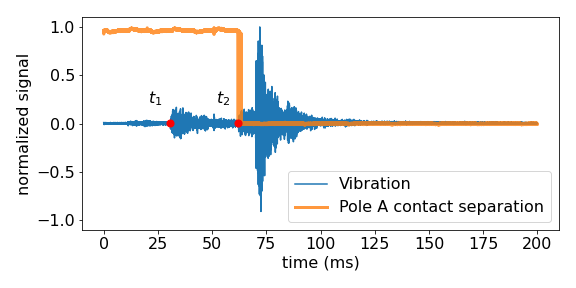}}
\caption{An example of a vibration and contact separation measurement signal. The key moments are indicated in the figure as $t_\mathrm{1}$ and $t_\mathrm{2}$.}
\label{fig:example_signal}
\end{figure}

\subsection{Vibration Signal and Key Moments}
\label{sec:vibration}
To measure vibration signals, an accelerometer 786A (Wilcoxon Sensing Technologies) is installed on the drive enclosure fixed with mounting stud next to pole B because this position is close to the drive and also easy to mount as shown in Figure~\ref{fig:experiment_setup}. It is a piezoelectric accelerometer with single direction, sensitivity of 100 mV/g, and frequency response range of 0.5Hz - 14kHz. This sensor is chosen because vibration signals from mechanical parts usually have a frequency lower than 10kHz~\cite{ukil2013monitoring}. The recording vibration direction is perpendicular to the ground. An example of vibration signals during close operation is shown in Figure~\ref{fig:example_signal}.

\begin{figure*}[htbp]
\centerline{\includegraphics[width=0.7\textwidth]{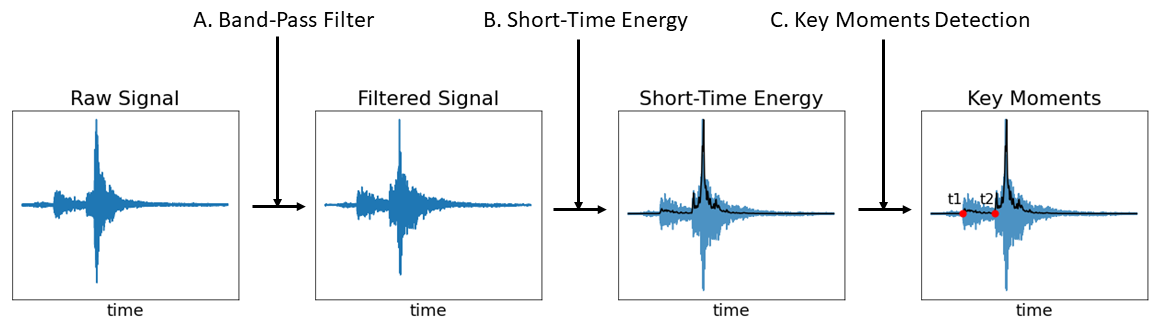}}
\caption{Flow chart of the proposed method for detecting key moments in close operation}
\label{fig:flowchart}
\end{figure*}

We define two key moments as specific time points where important events happen during the close operation. In Figure~\ref{fig:example_signal}, two key moments $t_\mathrm{1}$ and $t_\mathrm{2}$ (marked in red) can be distinguished. The first key moment $t_\mathrm{1}$ corresponds to the release of the latch. This induces increased vibration signals in the first segment of the signal. The second key moment $t_\mathrm{2}$ is the point where the moving contact touches the fixed contact, which corresponds also to the definition of the closing time~\cite{yang2019chaotic, sun2021remaining}. The start of the recording, time zero, in Figure~\ref{fig:example_signal} is defined when the trigger signal is sent from LabJack T7 to the VCB close coil. After receiving the trigger signal, the electromagnet of the VCB starts to energize and then the operating mechanism starts to move, resulting in a full close operation.



\subsection{Contact Separation Measurement}
\label{sec:contact_separation}
To measure the VCB opening and closing times, contact separation measurement is implemented as described in Section~\ref{sec:intro}. We connect a low direct current (DC) voltage of 9V between contacts. When the VCB is in its open position, a voltage difference can be measured between contacts. When it is in its close position, there is no voltage difference. To prevent a short circuit of the low-voltage source in close position, a 10k$\Omega$ series resistor is installed.

This measurement is the ground-truth of closing time because it enables to infer precisely when the contacts touch each other. An example signal of pole A is shown in color orange in Figure~\ref{fig:example_signal}. The normalized voltage signal drops from 1 to 0 at around 60ms during close operation. The time when the voltage signal first drops is the closing time. 



\section{Methods}
\label{sec:methods}
In this section, we present the proposed method for detecting key moments using vibration signals recorded by the accelerometer during close operation. A flow chart summarizing the proposed method is given in Figure~\ref{fig:flowchart}.

\subsection{Band-Pass Filter}
First, we filter the vibration signal using a band-pass filter because the accelerometer we used has a $\pm3$dB measurement range of 0.5Hz to 14kHz. The band-pass filter with the same lower and upper cutoff frequency of 0.5Hz and 14kHz is selected. 


\subsection{Short-Time Energy (STE)}
\label{sec:ste}
In speech applications, STE has demonstrated to be able to distinguish voiced and unvoiced parts of recording signals and, thus, achieve silence removal~\cite{jalil2013short, schafer1975digital}. In the field of CB, STE was also already used to characterise the vibration profile during closing operation~\cite{meng2006detection}. In this research, we propose to use the STE for detecting the closing time key moments in vibration signals. 

The signal energy $E$ for a discrete time signal $x(t)$ $\forall t \in \{0,1,...,T-1\}$ with $T$ the recording length can be found as: 

\begin{equation}
\label{eq:energy}
    E=\sum^{T-1}_{t=0}x^2(t)
\end{equation}

However, the vibration signal from the VCB is non-stationary and its characteristics change over time. We, therefore, use STE, which is commonly used for non-stationary signals. Similar to Equation~(\ref{eq:energy}), the STE is an energy measure. However, instead of taking the entire signal, we calculate the energy locally. 









The STE is defined as:
\begin{equation}
\label{eq:ste}
\begin{aligned}
    \mathrm{STE}(n) &= \sum_{t=-\infty}^{\infty}x^2(t)w^2(n-t)
\end{aligned}
\end{equation}
 
where $w(\bullet)$ is a window function. It determines the characteristics of the STE results. One commonly used window function is the Hamming window~\cite{blackman1958measurement} defined in Equation~(\ref{eq:hamming}) as $h(\bullet)$. The window length $W$ influences the STE results and needs to be specified by the user and $n$ is the sample number. The larger the window length $W$ is, the smoother the STE results are.


\begin{equation}
\label{eq:hamming}
    h(n)=
\begin{cases}
   0.54 - 0.46 \cos(\frac{2\pi n}{W-1}), &0\leq n \leq W-1  \\
    0,              & \text{otherwise}
\end{cases}
\end{equation}




\subsection{Key Moments Detection}


After computing the STE of the vibration signal, a decision rule needs to be defined that automatically determines the key moments $t_\mathrm{1}$ and $t_2$. We propose a detection algorithm with a simple procedure and few hyperparameters to make it generalizable to other types of CBs. It detects key moments based on the moving threshold and is explained in Algorithm~\ref{alg:closing_time_detection}.







This algorithm contains only four hyperparameters that need to be set by the user. By using different combinations of hyperparameters, different key moments can be detected. First, we specify the first two hyperparameters, a pre-defined detection time interval from $t_{\mathrm{start}}$ to $t_{\mathrm{end}}$. Generally, the information about the expected range of closing time is provided in the CB manual. Therefore,  it is possible to infer suitable $t_{\mathrm{start}}$ and $t_{\mathrm{end}}$ from it. Then, we select
$t_0$, which is the detection starting point, where $0 < t_\mathrm{0} < t_{\mathrm{end}}-t_{\mathrm{start}}$. We start the iteration over $t$ by calculating the mean $\mu$ and the standard deviation $\sigma$ of the signal between $t_{\mathrm{start}}$ and $t_{\mathrm{start}} + t$, where $t$ starts from $t_0$. In other words, the detection starts from $t_{\mathrm{start}}+t_\mathrm{0}$ and iterates over $t$ from $t_{\mathrm{start}}+t_\mathrm{0}$ to $t_{\mathrm{end}}$ by checking if the absolute difference between the signal at time $t$, $\text{Signal}(t)$, to the mean $\mu$ is larger than a pre-defined threshold which is set to three times standard deviation, $3\sigma$.

\begin{figure}[htbp]
\centerline{\includegraphics[width=70mm]{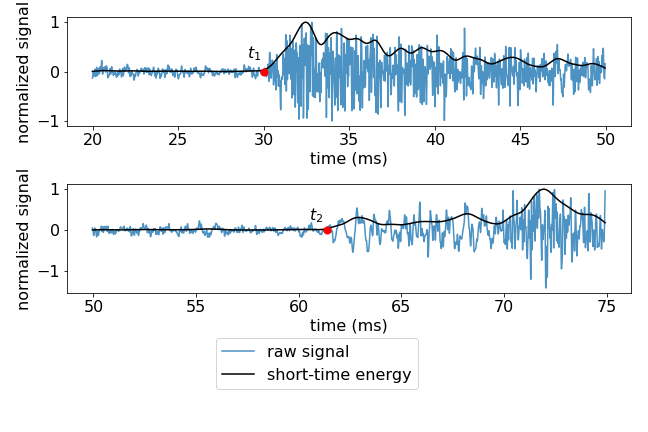}}
\caption{The key moments $t_\mathrm{1}$ and $t_\mathrm{2}$ detected from Algorithm~\ref{alg:closing_time_detection}}
\label{fig:key_point_t1_t2_example}
\end{figure}

There is always noise in the signal causing energy fluctuations through time. If we would directly take the first point above the threshold, the decision algorithm would not be robust. To make the decision rule more robust, we introduce another hyperparameter $L$, which is the tolerated number of points above the threshold. It defines how many consecutive points above the threshold are needed to be a valid detection. This makes the decision rule more robust to noise. If no point is above the threshold, the algorithm returns -1 as shown in step 17, indicating that no key moment is detected. An example of a vibration signal and the two detected key moments are shown in Figure~\ref{fig:key_point_t1_t2_example}.

\begin{algorithm}
\caption{Key Moments Detection Algorithm}\label{alg:closing_time_detection}
\begin{algorithmic}[1]

\State $\mathrm{Signal} \gets \mathrm{Signal}_{ t_{\mathrm{start}}:t_{\mathrm{end}}}$
\State $t \gets t_0$
\State $P \gets 0$
\State Calculate the mean $\mu$ and standard deviation $\sigma$ of $\mathrm{Signal}_{0:t}$
\For{$t \gets t_0$ to $t_{\mathrm{end}}-t_{\mathrm{start}}$}
\If{$|\mathrm{Signal}(t) - \mu| \geq 3\sigma$}
    \State $P \gets P+1$
    \State go to step 5
\Else
    \State go to step 4
\EndIf

\If{$P>L$}
    \State Key moment = $t-P$
    \State go to step 18
\EndIf
\EndFor
\State Key moment $\gets -1$\\
\Return Key moment
\end{algorithmic}
\end{algorithm}





\subsection{Change Point Detection}
\label{sec:change_point}
We compared our proposed detection algorithm with three selected change point detection algorithms from the literature~\cite{truong2020selective} as explained in Section~\ref{sec:literature}, including BS, window sliding, and bottom-up segmentation. For simplicity reasons, we only present the details of BS because it has the best performance among the three selected methods. 

The BS algorithm finds the change point at time $t_{cp}$ by minimising the following cost function:

\begin{equation}
\label{eq:binseg}
t_{cp} = \mathrm{argmin}_{1\leq t<T-1}c(x_{0:t})+c(x_{t:T})
\end{equation}

where $c(\bullet)$ is the cost function and $x_{0:t}$ is the signal from time $0$ to $t$, $T$ is the record length. One of the commonly used cost functions is the empirical mean $c_{L_2}$ as defined in Equation~\ref{eq:l2}, where $\bar{x}_{a,b}$ is the empirical mean of $x_{a:b}$ and $a, b$ are the starting and ending time.

\begin{equation}
\label{eq:l2}
c_{L_2}(x_{a:b}) = \sum^{b}_{t=a+1}\lVert x_t-\bar{x}_{a:b}\rVert^2_2
\end{equation}

\subsection{Hyperparameter Setting}

Hyperparameter setting for Algorithm~\ref{alg:closing_time_detection} is provided in Table~\ref{table:hyperparameters}. According to the VCB manual, we know that the closing time should be around 60ms. We also know that the first key moment $t_\mathrm{1}$ occurs before $t_\mathrm{2}$. Therefore, we select $t_0$ to be 10ms for both $t_\mathrm{1}$ and $t_\mathrm{2}$. Similarly, $t_{\mathrm{start}}$ is set to be 20ms and 50ms respectively for $t_\mathrm{1}$ and $t_\mathrm{2}$. For $t_{\mathrm{end}}$, we select 50ms for $t_\mathrm{1}$ and 75ms for $t_\mathrm{2}$. We set $L$ to be 300 points, which means that the signal is allowed to be above threshold for at most 300 consecutive points or 1ms when the sampling rate is 300kHz as in our case.  

In addition, for STE, we use the Hamming window with a window length $W$ of 600. For our application, the detection algorithm is not sensitive to the window length. We use zero padding. Therefore, the output signal has the same length as the input signal.

For change point detection, a BS algorithm is applied to the energy of the signal as defined in Equation~\ref{eq:energy}. The pre-defined detection range for key moment $t_\mathrm{cp}$ is set to be between 60ms to 85ms and the cost function is the empirical mean in Equation~\ref{eq:l2}.



\begin{table}[htbp]
\caption{Hyperparameter setting for Algorithm~\ref{alg:closing_time_detection}}
\begin{center}
\begin{tabular}{|c|c|c|c|c|}
\hline
key moment & \textbf{$t_{\mathrm{start}}$} (ms)& \textbf{$t_{\mathrm{end}}$} (ms)& \textbf{$t_0$} (ms)& \textbf{$L$} (\# point(s))\\
\hline\hline
$t_1$ & 20& 50& 10& 300  \\
\hline
$t_2$ & 50& 75& 10& 300  \\
\hline
\end{tabular}
\label{table:hyperparameters}
\end{center}
\end{table}





\section{Results}
\label{sec:results}

\subsection{Closing Time Evolution}
\begin{figure}[htbp]
\centerline{\includegraphics[width=60mm]{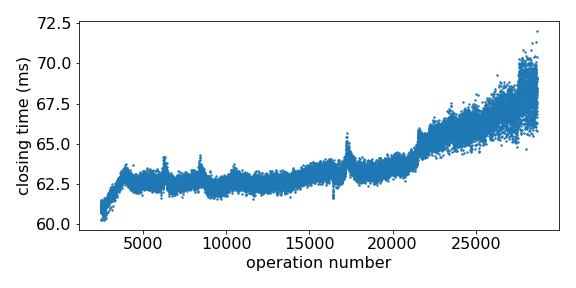}}
\caption{Pole A closing time $t_\mathrm{c}$ from the contact separation measurement}
\label{fig:closing_time}
\end{figure}

Figure~\ref{fig:closing_time} shows the closing time measured from the contact separation measurement of pole A. It increases when the operation number increases and starts from around 60ms as described in the VCB manual to more than 70ms by the end of life. Pole B and C show a similar trend. Therefore, only the closing time of pole A is shown here for simplicity.

\subsection{Key Moments Detection}
\begin{figure}[htbp]
\centerline{\includegraphics[width=70mm]{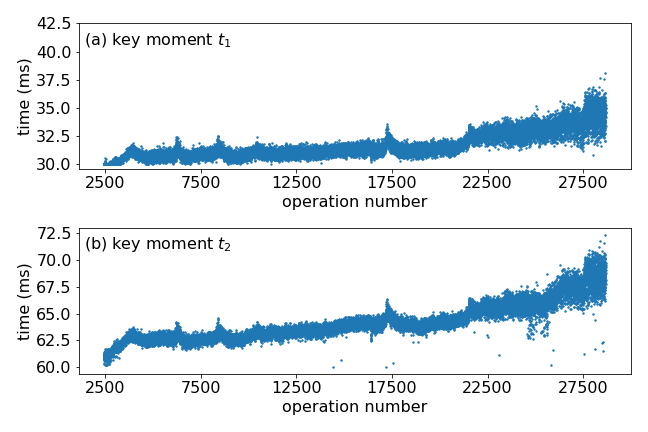}}
\caption{Key moments (a) $t_1$ and (b) $t_2$ detected using Algorithm~\ref{alg:closing_time_detection} over the entire VCB lifetime}
\label{fig:key_point_t1_t2}
\end{figure}

Figure~\ref{fig:key_point_t1_t2} shows the key moments $t_\mathrm{1}$ and $t_\mathrm{2}$ over the entire VCB lifetime detected by Algorithm~\ref{alg:closing_time_detection}. The overall increasing trends are similar in the course of the whole lifetime for both $t_\mathrm{1}$ and $t_\mathrm{2}$, and also to the closing time $t_\mathrm{c}$ in Figure~\ref{fig:closing_time} from the contact separation measurements. 

\begin{figure}[htbp]
\centerline{\includegraphics[width=60mm]{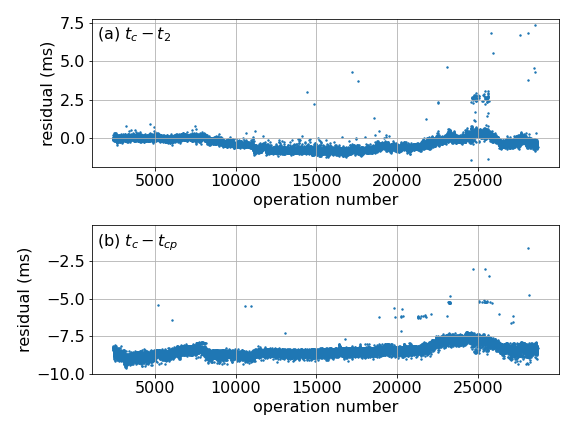}}
\caption{The residual (a) $t_\mathrm{c} - t_\mathrm{2}$ and (b) $t_\mathrm{c} - t_\mathrm{cp}$ over the entire VCB lifetime}
\label{fig:residual}
\end{figure}

In Figure~\ref{fig:residual}(a), we show the residual between the closing time from the contact separation measurement and our proposed detection algorithm $t_\mathrm{c} - t_\mathrm{2}$. A discrepancy appears at $i=\mathrm{25000}$ where our detection algorithm underestimates the closing time, because the vibration patterns change within this region. In Figure~\ref{fig:residual}(b), the residual between the closing time and the BS results $t_{\mathrm{c}} - t_{\mathrm{cp}}$ are displayed. The change points $t_\mathrm{cp}$ have a delay of ca. 8ms compared to the closing time $t_{\mathrm{c}}$. However, no outliers are detected in this case. This indicates that there was probably no technical underlying reason for these outliers. 

We further quantify the performance using the root mean squared error (RMSE) as defined in Equation~(\ref{eq:mse}) of the detected key moment $t_\mathrm{detected}$  and the closing time $t_c$ derived from the contact separation measurement. The operation number $i$ goes from $i_{\mathrm{start}}=\mathrm{2442}$ to $i_{\mathrm{end}}=\mathrm{28684}$. For our proposed algorithm, $t_\mathrm{detected}$ is $t_\mathrm{2}$ and for the BS, $t_\mathrm{detected}$ is $t_\mathrm{cp}$.

\begin{equation}
\label{eq:mse}
    \mathrm{RMSE} = \sqrt{\frac{1}{i_{\mathrm{end}}-i_{\mathrm{start}}+1}\sum_{i=i_{\mathrm{start}}}^{i_{\mathrm{end}}} (t_{\mathrm{c},i} - t_{\mathrm{detected}, i})^2}
\end{equation}

For the proposed algorithm, the RMSE is 0.550ms for the entire VCB life and for the BS algorithm the RMSE is 8.438ms. If we remove the delay by taking the mean of the first 5000 operations as the reference value of the delay, the RMSE of the BS algorithm reduces to 0.466ms. We take the first 5000 operations because the closing time in the early stage is more stable. Even if the RMSE is lower for the BS algorithm after removing the delay, we would need to know that there is a delay in the first place. However, for this, we would need the ground truth information from the intrusive monitoring. 


\subsection{Key Moments Analysis}
\begin{figure}[htbp]
\centerline{\includegraphics[width=60mm]{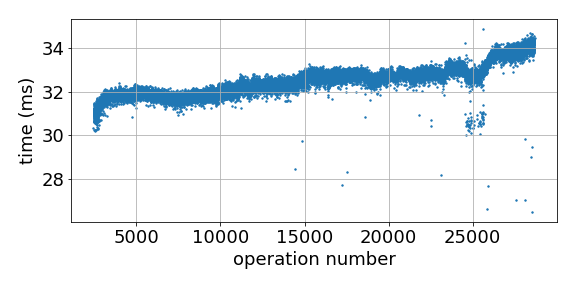}}
\caption{The time interval $t_\mathrm{2} - t_\mathrm{1}$. It shows that not only $t_\mathrm{1}$ and $t_\mathrm{2}$ are increasing but also $t_\mathrm{2} - t_\mathrm{1}$ is increasing}
\label{fig:time_interval_key_points}
\end{figure}

The closing time $t_\mathrm{c}$ increases when the operation number increases according to the contact separation measurements. To further understand where this increase in time comes from, we analyze the time difference between key moments $t_\mathrm{2}-t_\mathrm{1}$. This corresponds to the contact moving time, starting from the latch movement to the time when the contacts touch each other. This time interval also increases as the operation number increases and is plotted in Figure~\ref{fig:time_interval_key_points}. This indicates that the whole operating mechanism moving time increases. In other words, the increase in closing time is caused not only by the increase of the latch time $t_\mathrm{1}$ but also by the increase of the operating mechanism moving time $t_\mathrm{2}-t_\mathrm{1}$. This information is not available from the contact separation measurements. 



%


\section{Discussion}
\label{sec:discussion}
\textbf{Closing Time:} The increasing trend of closing time indicates the wearing of the VCB operating mechanisms due to the continuous operations. By the end of the VCB's life there is a region after $i=\mathrm{27000}$ where closing time increases even above 70ms, which is about 10ms more than the upper limit of the expected closing time specified in the VCB manual. This implies that even though the VCB is still able to perform mechanical close operations, it may not function properly to carry or interrupt the current. The trend of the closing time $t_\mathrm{c}$ in Figure~\ref{fig:closing_time} can be divided in three stages based on their characteristics: initiation ($i<\mathrm{5000}$), stationary ($\mathrm{5000}<i<\mathrm{20000}$), and wearing stage ($i>\mathrm{20000}$). The first initiation stage has low closing time but high standard deviation. The stationary stage has stable mean and low standard deviation. In the final wearing stage, both the mean and the standard deviation increase.

 


Besides the overall increasing trend in closing time, we can observe some anomalies: 1) two peaks appear between the operation number 5000 and 10000 and 2) one peak occurs between the operation number 15000 and 20000. The causes of these peaks could not be explained by domain experts and the underlying reasons for these peaks could not be clarified.


\textbf{Key Moments Detection}: Our proposed key moments detection algorithm is able to detect $t_\mathrm{2}$ accurately without any delay, which is verified by the closing time from the contact separation measurements. The BS algorithm is also able to detect the closing time but with a delay of approximately 8ms. The closing time increases because both latch time $t_\mathrm{1}$ and the operating mechanism time $t_\mathrm{2} - t_\mathrm{1}$ increase. This implies that the wearing of the VCB has an impact on the latch and also on the operating mechanisms. In case only $t_\mathrm{1}$ increases but not $t_\mathrm{2}-t_\mathrm{1}$, it could be used as an indicator for faults that are only impacting the latch.






\section{Conclusions and Outlook}
\label{sec:conclusions}
In this work, we proposed a key moments detection algorithm for identifying key moments during close operation from vibration signals during an accelerated VCB life testing without load. The closing time inferred from the vibration signals can be used for non-intrusive VCB condition monitoring. Our proposed method is able to perform continuous condition monitoring without the need to disconnect VCB from the grid. With the detected key moments, it is even possible to distinguish between the closing time and the latch initiation time. Our proposed algorithm is able to detect the closing time precisely without any delay. The alternative BS algorithm shows a substantial delay in the detection. 

The collected  dataset from the accelerated lifetime testing  provides information about the evolution of the VCB condition. This is the first time that a full trajectory of time to failure has been continuously monitored. In the conducted experiment, the VCB was able to operate more than 26000 open-close operations until the friction in the operating mechanism increased to a level that prevents it from further operation. The dataset was made publicly available.


This work opens many research directions. For example, in substations, CBs are always operated with current load. Therefore, the vibration patterns could be different from those without load. The monitoring should, thus, be extended to load conditions as well. Furthermore, we observed a continuous degradation. However, there are possible sudden or evolving failures, resulting in different fault patterns and fault evolution trends. 
Moreover, it is important to emphasize that the vibration signals contain more information than the latch and the closing time. Evaluating this additional information on the health condition and also evaluating how the vibration signals may be affected by other potential fault types is left for future research. Finally, the application of our detection method on different types of CBs is left for future research. 






\bibliographystyle{IEEEtran} 
\bibliography{reference}

\begin{thebibliography}{10}
\providecommand{\url}[1]{#1}
\csname url@samestyle\endcsname
\providecommand{\newblock}{\relax}
\providecommand{\bibinfo}[2]{#2}
\providecommand{\BIBentrySTDinterwordspacing}{\spaceskip=0pt\relax}
\providecommand{\BIBentryALTinterwordstretchfactor}{4}
\providecommand{\BIBentryALTinterwordspacing}{\spaceskip=\fontdimen2\font plus
\BIBentryALTinterwordstretchfactor\fontdimen3\font minus
  \fontdimen4\font\relax}
\providecommand{\BIBforeignlanguage}[2]{{%
\expandafter\ifx\csname l@#1\endcsname\relax
\typeout{** WARNING: IEEEtran.bst: No hyphenation pattern has been}%
\typeout{** loaded for the language `#1'. Using the pattern for}%
\typeout{** the default language instead.}%
\else
\language=\csname l@#1\endcsname
\fi
#2}}
\providecommand{\BIBdecl}{\relax}
\BIBdecl

\bibitem{ieee1992ieee}
I.~P.~S. Committee \emph{et~al.}, ``Ieee standard definitions for power
  switchgear,'' \emph{IEEE Std. C}, vol.~37, pp. 100--1992, 1992.

\bibitem{janssen2013international}
A.~Janssen, D.~Makareinis, and C.-E. S{\"o}lver, ``International surveys on
  circuit-breaker reliability data for substation and system studies,''
  \emph{IEEE Transactions on Power Delivery}, vol.~29, no.~2, pp. 808--814,
  2013.

\bibitem{Carvalho2012CIGRTBP1}
A.~Carvalho, M.~L. Cormenzana, H.~Furuta, W.~Grieshaber, A.~Hyrczak,
  D.~Kopejtkova, J.~Krone, M.~Kudoke, D.~Makareinis, J.~Martins, K.~Mestrovic,
  I.~Ohno, J.~Ostlund, K.~Park, J.~Patel, C.~Protze, M.~Runde, J.~Schmid,
  J.~Skog, C.~Solver, B.~Sweeney, and F.~Waite, ``Cigre technical brochure no.
  509, 510: Final report of the 2004 – 2007 international enquiry on
  reliability of high voltage equipment, part 1 - summary and general matters,
  part 2 - reliability of high voltage sf6 circuit breakers,'' 2012.

\bibitem{razi2020condition}
A.~A. Razi-Kazemi and K.~Niayesh, ``Condition monitoring of high voltage
  circuit breakers: Past to future,'' \emph{IEEE Transactions on Power
  Delivery}, vol.~36, no.~2, pp. 740--750, 2020.

\bibitem{razi2015applicability}
A.~Razi-Kazemi, ``Applicability of auxiliary contacts in circuit breaker online
  condition assessment,'' \emph{Electric Power Systems Research}, vol. 128, pp.
  53--59, 2015.

\bibitem{meng2006detection}
Y.~Meng, S.~Jia, Z.~Shi, and M.~Rong, ``The detection of the closing moments of
  a vacuum circuit breaker by vibration analysis,'' \emph{IEEE transactions on
  power delivery}, vol.~21, no.~2, pp. 652--658, 2006.

\bibitem{hoidalen2005continuous}
H.~Hoidalen and M.~Runde, ``Continuous monitoring of circuit breakers using
  vibration analysis,'' \emph{IEEE Transactions on Power Delivery}, vol.~20,
  no.~4, pp. 2458--2465, 2005.

\bibitem{yang2019chaotic}
Q.~Yang, J.~Ruan, Z.~Zhuang, and D.~Huang, ``Chaotic analysis and feature
  extraction of vibration signals from power circuit breakers,'' \emph{IEEE
  Transactions on Power Delivery}, vol.~35, no.~3, pp. 1124--1135, 2019.

\bibitem{dou2018application}
L.~Dou, S.~Wan, and C.~Zhan, ``Application of multiscale entropy in mechanical
  fault diagnosis of high voltage circuit breaker,'' \emph{Entropy}, vol.~20,
  no.~5, p. 325, 2018.

\bibitem{yang2019new}
Q.~Yang, J.~Ruan, Z.~Zhuang, D.~Huang, and Z.~Qiu, ``A new vibration analysis
  approach for detecting mechanical anomalies on power circuit breakers,''
  \emph{IEEE Access}, vol.~7, pp. 14\,070--14\,080, 2019.

\bibitem{ma2018intelligent}
S.~Ma, M.~Chen, J.~Wu, Y.~Wang, B.~Jia, and Y.~Jiang, ``Intelligent fault
  diagnosis of hvcb with feature space optimization-based random forest,''
  \emph{Sensors}, vol.~18, no.~4, p. 1221, 2018.

\bibitem{ukil2013monitoring}
A.~Ukil, M.~Zlatanski, and M.~Hochlehnert, ``Monitoring of hv generator circuit
  breaker contact ablation based on acoustic emission,'' \emph{IEEE
  Transactions on Instrumentation and Measurement}, vol.~62, no.~10, pp.
  2683--2693, 2013.

\bibitem{johal2008coil}
H.~Johal and M.~Mousavi, ``Coil current analysis method for predictive
  maintenance of circuit breakers,'' in \emph{2008 IEEE/PES Transmission and
  Distribution Conference and Exposition}.\hskip 1em plus 0.5em minus
  0.4em\relax IEEE, 2008, pp. 1--7.

\bibitem{razi2016circuit}
A.~A. Razi-Kazemi, ``Circuit breaker condition assessment through a
  fuzzy-probabilistic analysis of actuating coil's current,'' \emph{IET
  Generation, Transmission \& Distribution}, vol.~10, no.~1, pp. 48--56, 2016.

\bibitem{liu2018prediction}
Y.~Liu, G.~Zhang, H.~Qin, Y.~Geng, J.~Wang, J.~Yang, and K.~Zhao, ``Prediction
  of the dynamic contact resistance of circuit breaker based on the kernel
  partial least squares,'' \emph{IET Generation, Transmission \& Distribution},
  vol.~12, no.~8, pp. 1815--1821, 2018.

\bibitem{rusek2008timings}
B.~Rusek, G.~Balzer, M.~Holstein, and M.-S. Claessens, ``Timings of high
  voltage circuit-breaker,'' \emph{Electric power systems research}, vol.~78,
  no.~12, pp. 2011--2016, 2008.

\bibitem{truong2020selective}
C.~Truong, L.~Oudre, and N.~Vayatis, ``Selective review of offline change point
  detection methods,'' \emph{Signal Processing}, vol. 167, p. 107299, 2020.

\bibitem{wang2018hybrid}
B.~Wang, Y.~Lei, N.~Li, and N.~Li, ``A hybrid prognostics approach for
  estimating remaining useful life of rolling element bearings,'' \emph{IEEE
  Transactions on Reliability}, vol.~69, no.~1, pp. 401--412, 2018.

\bibitem{arias2021aircraft}
M.~Arias~Chao, C.~Kulkarni, K.~Goebel, and O.~Fink, ``Aircraft engine
  run-to-failure dataset under real flight conditions for prognostics and
  diagnostics,'' \emph{Data}, vol.~6, no.~1, p.~5, 2021.

\bibitem{obarcanin2021}
K.~Obarcanin, D.~Skulj, and B.~Lacevic, ``High voltage circuit breaker
  vibration signature indices evaluation for condition assessment,'' \emph{B\&H
  Electrical Engineering}, vol.~15, pp. 82--88, 2021.

\bibitem{vcb_dataset2022}
\BIBentryALTinterwordspacing
C.-C. Hsu, M.~Muratovic, and C.~Franck, ``Run-to-failure vacuum circuit breaker
  mechanical test dataset,'' 2022. [Online]. Available:
  \url{https://doi.org/20.500.11850/544221}
\BIBentrySTDinterwordspacing

\bibitem{sun2021remaining}
S.~Sun, Z.~Wen, T.~Du, J.~Wang, Y.~Tang, and H.~Gao, ``Remaining life
  prediction of conventional low-voltage circuit breaker contact system based
  on effective vibration signal segment detection and mccae-lstm,'' \emph{IEEE
  Sensors Journal}, vol.~21, no.~19, pp. 21\,862--21\,871, 2021.

\bibitem{jalil2013short}
M.~Jalil, F.~A. Butt, and A.~Malik, ``Short-time energy, magnitude, zero
  crossing rate and autocorrelation measurement for discriminating voiced and
  unvoiced segments of speech signals,'' in \emph{2013 The international
  conference on technological advances in electrical, electronics and computer
  engineering (TAEECE)}.\hskip 1em plus 0.5em minus 0.4em\relax IEEE, 2013, pp.
  208--212.

\bibitem{schafer1975digital}
R.~W. Schafer and L.~R. Rabiner, ``Digital representations of speech signals,''
  \emph{Proceedings of the IEEE}, vol.~63, no.~4, pp. 662--677, 1975.

\bibitem{blackman1958measurement}
R.~B. Blackman and J.~W. Tukey, ``The measurement of power spectra from the
  point of view of communications engineering—part i,'' \emph{Bell System
  Technical Journal}, vol.~37, no.~1, pp. 185--282, 1958.

\end{thebibliography}

\end{document}